\documentclass[aps, twocolumn, superscriptaddress, showpacs, longbibliography]{revtex4-1}

\usepackage{amsmath, amsfonts, amssymb, amsthm}
\usepackage{hyperref}
\usepackage{bbm}
\usepackage{physics}
\usepackage{tabularx}
\usepackage{tikz}
\usepackage{tensor}
\usepackage{usefulnotations}
\usepackage{multirow}

\newcommand{\Sbra}[1]{\tensor[_{\text{\tiny S}}]{\bra{#1}}{}}
\newcommand{\Sket}[1]{\tensor{\ket{#1}}{_{\text{\tiny S}}}}

\newcommand{\Hket}[1]{\tensor{\ket{#1}}{_{\text{\tiny H}}}}

\newcommand{\Vbra}[1]{\tensor[_{\text{\tiny V}}]{\bra{#1}}{}}
\newcommand{\Vket}[1]{\tensor{\ket{#1}}{_{\text{\tiny V}}}}
\newcommand{\Vbraket}[2]{\tensor[_{\text{\tiny V}}]{\braket{#1}{#2}}{_{\text{\tiny V}}}}
\newcommand{\HLbra}[1]{\tensor[_{\text{\tiny HL}}]{\bra{#1}}{}}
\newcommand{\HLket}[1]{\tensor{\ket{#1}}{_{\text{\tiny HL}}}}

\newcommand{\SBra}[1]{\tensor[_{\text{\tiny S}}]{\left\llangle #1 \right|}{}}
\newcommand{\SKet}[1]{\tensor{\left| #1 \right\rrangle}{_{\text{\tiny S}}}}
\newcommand{\SBraket}[2]{\tensor[_{\text{\tiny S}}]{\left\llangle #1 \vphantom{#2} \right|\kern-0.6ex\left. #2 \vphantom{#1}\right\rrangle}{_{\text{\tiny S}}}}
\newcommand{\HBra}[1]{\tensor[_{\text{\tiny H}}]{\left\llangle #1 \right|}{}}
\newcommand{\HKet}[1]{\tensor{\left| #1 \right\rrangle}{_{\text{\tiny H}}}}
\newcommand{\HBraket}[2]{\tensor[_{\text{\tiny H}}]{\left\llangle #1 \vphantom{#2} \right|\kern-0.6ex\left. #2 \vphantom{#1}\right\rrangle}{_{\text{\tiny H}}}}
\newcommand{\ind}[2]{#1_\text{\tiny #2}}

\begin{document}
	\title{Heisenberg and Heisenberg-Like Representations via Hilbert Space Bundle Geometry in the Non-Hermitian Regime}

	\author{Chia-Yi Ju}
	\affiliation{Department of Physics, National Sun Yat-sen University, Kaohsiung 80424, Taiwan}
	\affiliation{Center for Theoretical and Computational Physics, National Sun Yat-sen University, Kaohsiung 80424, Taiwan}
	\affiliation{Physics Division, National Center for Theoretical Sciences, Taipei 106319, Taiwan}
	\author{Adam Miranowicz}
	\affiliation{Institute of Spintronics and Quantum Information, Faculty of Physics and Astronomy, Adam Mickiewicz University, 61-614 Pozna\'n, Poland}
	\affiliation{Center for Quantum Computing, RIKEN, Wakoshi, Saitama 351-0198, Japan}
	\author{Jacob Barnett}
	\affiliation{Basque Center for Applied Mathematics, Alameda de Mazarredo 14, Bilbao 48009, Bizkaia, Spain}
	\author{Guang-Yin Chen}
	\email{gychen@phys.nchu.edu.tw}
	\affiliation{Department of Physics, National Chung Hsing University, Taichung 40227, Taiwan}
	\affiliation{Physics Division, National Center for Theoretical Sciences, Taipei 106319, Taiwan}
	\author{Franco Nori}
	\affiliation{Center for Quantum Computing, RIKEN, Wakoshi, Saitama 351-0198, Japan}
	\affiliation{Department of Physics, University of Michigan, Ann Arbor, Michigan 48109-1040, USA}

	\date{\today}

	\begin{abstract}
		The equivalence between the Schr\"{o}dinger and Heisenberg representations is a cornerstone of quantum mechanics. However, this relationship remains unclear in the non-Hermitian regime, particularly when the Hamiltonian is time-dependent. In this study, we address this gap by establishing the connection between the two representations, incorporating the metric of the Hilbert space bundle. We not only demonstrate the consistency between the Schr\"{o}dinger and Heisenberg representations but also present a Heisenberg-like representation grounded in the generalized vielbein formalism, which provides a clear and intuitive geometric interpretation. Unlike the standard Heisenberg representation, where the metric of the Hilbert space is encoded solely in the dual states, the Heisenberg-like representation distributes the metric information between both the states and the dual states. Despite this distinction, it retains the same Heisenberg equation of motion for operators. Within this formalism, the Hamiltonian is replaced by a Hermitian counterpart, while the ``non-Hermiticity'' is transferred to the operators. Moreover, this approach extends to regimes with a dynamical metric (beyond the pseudo-Hermitian framework) and to systems governed by time-dependent Hamiltonians.
	\end{abstract}

	\maketitle
	\newpage

	\section{Introduction}
		Quantum theory is widely recognized as the most accurate and comprehensive physical model for understanding how nature operates. To carry out theoretical calculations, a specific representation is needed to mathematically describe the states and observables of a quantum system. Within the framework of quantum theory, numerous equivalent yet seemingly distinct representations exist. Among these, the Schr\"{o}dinger and Heisenberg representations are arguably the most well-known.

		The advent of parity-time ($\mathcal{PT}$) symmetric non-Hermitian quantum mechanics~\cite{Bender1998, Bender2004, Bender2007}, along with various other frameworks for non-Hermitian quantum mechanics~\cite{Mostafazadeh2003, Mostafazadeh2004, Mostafazadeh2010, Brody2013, Ju2019}, raises a critical question: do the Schr\"{o}dinger and Heisenberg representations remain equivalent in the non-Hermitian regime?

		Specifically, the conventional transformation from the Schr\"odinger representation to the Heisenberg representation leads to several paradoxes. In Sec.~\ref{Sec:ConvHR}, we highlight discrepancies such as conflicting predictions of measurement outcomes and violations of canonical commutation relations. Given the increasing prominence of non-Hermitian quantum systems~\cite{Hodaei2017, Chen2017, Quijandria2018, Stefano2019, Ashida2020, Znojil2021, Tzeng2021, Tu2022, Arkhipov2023, Tzeng2023, Ju2024, Znojil2024a, Ju2024a, Yang2024, Bao2024}, resolving these inconsistencies has become a critical issue.

		While numerous studies~\cite{Bagarello2012, Znojil2015, Miao2016, Bagarello2022} have aimed to reconcile the discrepancies between the two representations, most are limited to the quasi-Hermitian regime, apply only to specific types of operators, or rely on switching between Hilbert spaces to ``Hermitize'' the system~\cite{Dyson1956, Znojil2024}. Furthermore, few (if any) have addressed cases where the Hamiltonian is time-dependent.

		In this work, we introduce a natural transformation between the Schr\"{o}dinger and Heisenberg representations (see Table~\ref{Table:Comparison}) by incorporating the metric of the Hilbert space bundle~\cite{Ju2019}. Additionally, we propose a Heisenberg-like representation that makes the geometric information implicit via the generalized vielbein formalism~\cite{Ju2022}. These formalisms inherently extend beyond the pseudo-Hermitian regime, accommodates a time-dependent metric (including regimes with exceptional points (EPs)~\cite{Kato1995, Heiss2004}), applies to all operator types and effectively handles time-dependent Hamiltonians. Consequently, taking the geometry of the Hilbert space into account not only resolves the discrepancies between the Schr\"{o}dinger and Heisenberg representations but also provides an alternative representation in which both the time dependence and the ``non-Hermiticity'' are carried by the operators.

		\begin{table*}
			\renewcommand{\arraystretch}{1.5}
			\begin{tabular}{|c|c|c|}
				\hline
				& Schr\"{o}dinger representation & Heisenberg representation\\
				\hline
				Dynamics & $\displaystyle \left[\partial_t + i \ind{H}{S}(t)\right] \Sket{\psi(t)} = 0$ & $\displaystyle \frac{d}{dt}\ind{\mathcal{O}}{H}(t) = i \big[ \ind{H}{H}(t), \ind{\mathcal{O}}{H}(t) \big] + \big[ \partial_t \mathcal{O} (t) \big]_\text{\tiny H}$ \vphantom{$\Bigg[$}\\
				\hline
				Time-evolution operators & \multicolumn{2}{c|}{$\displaystyle \partial_t U(t, t_0) = - i \ind{H}{S} (t) U(t, t_0)$}\\
				\hline
				States & \begin{tabular}{rl} $\displaystyle \SKet{\psi(t)}$ & $\displaystyle = \Sket{\psi(t)} = U(t, t_0) \Sket{\psi(t_0)}$\\ & $\displaystyle = U(t, t_0) \Hket{\psi(t_0)}$\end{tabular} & $\displaystyle \HKet{\psi} = \Hket{\psi} = \Sket{\psi(t_0)}$\\
				\hline
				Dual states & \begin{tabular}{rl}$\displaystyle \SBra{\psi(t)}$ & $\displaystyle =\Sbra{\psi(t)} G(t) = \Sbra{\psi(t_0)} G(t_0) U^{-1}(t, t_0)$\\ & $\displaystyle = \HBra{\psi} U^{-1}(t, t_0)$\end{tabular} & $\displaystyle \HBra{\psi} = \SBra{\psi(t_0)} = \Sbra{\psi(t_0)} G(t_0)$\\
				\hline
				Observables & $\displaystyle \ind{\mathcal{O}}{S}(t) = U(t, t_0) \ind{\mathcal{O}}{H}(t) U^{-1}(t, t_0)$ & $\displaystyle \ind{\mathcal{O}}{H}(t) = U^{-1}(t, t_0) \ind{\mathcal{O}}{S}(t) U(t, t_0)$\\
				\hline
			\end{tabular}
			\caption{Relationships between states, dual states, and observables, along with their governing equations of motion in the Schrödinger and Heisenberg representations in the non-Hermitian regime. In the Hermitian regime, these representations simplify to the standard forms under the trivial metric $G(t) = \mathbbm{1}$ and $U^{-1}(t, t_0) = U^\dagger(t, t_0)$.}
			\label{Table:Comparison}
		\end{table*}

	\section{The Conventional Heisenberg Representation \label{Sec:ConvHR}}

		We start by briefly reviewing the relationship between the Schr\"{o}dinger and Heisenberg representations in quantum mechanics.

		In the Schr\"{o}dinger representation, the states in a quantum system described by the Hamiltonian $\ind{H}{S} (t)$ (where the subscript ``S'' stands for the Schr\"{o}dinger representation), evolve according to
		\begin{equation}
			i \partial_t \Sket{\psi (t)} = \ind{H}{S} (t) \Sket{\psi (t)}. \label{OriginalSchroedingerEq}
		\end{equation}

		Moreover, the expectation value of an observable $\mathcal{O}$ is
		\begin{equation}
			\left< \mathcal{O} \right> (t) = \Sbra{\psi (t)} \ind{\mathcal{O}}{S} (t) \Sket{\psi (t)},
		\end{equation}
		where $\ind{\mathcal{O}}{S} (t)$ is the corresponding operator in the Schr\"{o}dinger representation.

		Rather than using Eq.~\eqref{OriginalSchroedingerEq}, the time-evolution of a state can also be determined by the time-evolution operator $U (t, t_0)$, which satisfies:
		\begin{align}
			& \partial_t U (t, t_0) = - i \ind{H}{S} (t) U (t, t_0),\\
			& U (t_0, t_0) = \mathbbm{1}.
		\end{align}
		To be more specific,
		\begin{equation}
			\Sket{\psi (t)} = U (t, t_0) \Sket{\psi (t_0)}
		\end{equation}
		also evolves according to the Schr\"{o}dinger equation in Eq.~\eqref{OriginalSchroedingerEq}.

		In contrast, in the Heisenberg representation, the quantum states do not depend on time; instead, while the observables carry all the time dependence. Therefore, the time-evolution operator acts as the connection between the Heisenberg and Schr\"{o}dinger representations.

		Explicitly, the relationship between operators in the Schr\"{o}dinger and Heisenberg representations is given by
		\begin{equation}
			\ind{\mathcal{O}}{H} (t) = U^\dagger (t, t_0) \ind{\mathcal{O}}{S} (t) U (t, t_0), \label{HeisengbergAndSchroedingerOperator}
		\end{equation}
		while leaving the state fixed, i.e.,
		\begin{equation}
			\Hket{\psi} = \Sket{\psi (t_0)},
		\end{equation}
		where $t_0$ can be chosen arbitrarily for convenience, and the subscript ``H'' stands for the Heisenberg representation.

		Since $U^\dagger (t, t_0) = U^{-1} (t, t_0)$ for $\ind{H}{S}^\dagger = \ind{H}{S}$, it is clear that $\ind{\mathcal{O}}{H} (t)$ is a similarity transformation of $\ind{\mathcal{O}}{S} (t)$, and they share the same set of eigenvalues. In other words, not only the expectation values but also their possible measurement outcomes are the same.

		Moreover, if ${\lbrace (q_{\text{\tiny S}i}, p_{\text{\tiny S}i}) \rbrace}$ is a set of canonical conjugate quantity pairs, the commutation relations between them in the Schr\"{o}dinger representation are:
		\begin{align}
			& \big[ q_{\text{\tiny S}i}, q_{\text{\tiny S}j} \big] = 0,\\
			& \big[ p_{\text{\tiny S}i}, p_{\text{\tiny S}j} \big] = 0,\\
			& \big[ q_{\text{\tiny S}i}, p_{\text{\tiny S}j} \big] = i \delta_{ij},
		\end{align}
		where $\delta$ denotes the Kronecker delta.

		By applying the property $U^\dagger (t, t_0) = U^{-1} (t, t_0)$ once again, we find that the commutation relations in the Heisenberg representation remain unchanged. Specifically,
		\begin{align}
			\big[ q_{\text{\tiny H}i} (t), q_{\text{\tiny H}j} (t) \big] & = U^\dagger (t, t_0) \big[ q_{\text{\tiny S}i}, q_{\text{\tiny S}j} \big] U (t, t_0) = 0,\label{CommutationRelations1}\\
			\big[ p_{\text{\tiny H}i} (t), p_{\text{\tiny H}j} (t) \big] & = U^\dagger (t, t_0) \big[ p_{\text{\tiny S}i}, p_{\text{\tiny S}j} \big] U (t, t_0) = 0,\\
			\begin{split}
				\big[ q_{\text{\tiny H}i} (t), p_{\text{\tiny H}j} (t) \big] & = U^\dagger (t, t_0) \big[ q_{\text{\tiny S}i}, p_{\text{\tiny S}j} \big] U (t, t_0)\\
				& = U^\dagger (t, t_0) i \delta_{ij} U (t, t_0) = i \delta_{ij}.
			\end{split}\label{CommutationRelations3}
		\end{align}

		This outcome is, in fact, expected: The commutation relations must remain valid after transitioning to the Heisenberg representation. Failure to preserve these relations would not only violate the uncertainty principle, one of the most important and fundamental properties of quantum mechanics, but also undermine the canonical quantization framework.

		More specifically, the equivalence comes from the fact that the mapping between the Schr\"{o}dinger and Heisenberg representations is an invertible automorphism. To see this, let $\tau_{t, t_0}$ be a linear map defined as
		\begin{align}
			\tau_{t, t_0}(\mathcal{O}) = U^\dag(t, t_0) \mathcal{O} U(t, t_0),
		\end{align}
		where $U(t, t_0)$ is the evolution operator. Since $U^\dag(t, t_0) = U^{-1}(t, t_0)$, the structure of the operator is preserved, namely,
		\begin{align}
			& \tau_{t, t_0}(\mathcal{O}_\text{A} \mathcal{O}_\text{B}) = U^\dag(t, t_0) \mathcal{O}_\text{A} \mathcal{O}_\text{B} U(t, t_0)\\
			& = U^\dag(t, t_0) \mathcal{O}_\text{A} U(t, t_0) U^\dag(t, t_0) \mathcal{O}_\text{B} U(t, t_0)\\
			& = \tau_{t, t_0}(\mathcal{O}_\text{A}) \tau_{t, t_0}(\mathcal{O}_\text{B}).
		\end{align}

		Nevertheless, when the Hamiltonian is non-Hermitian, $\tau_{t, t_0}$ fails to be an algebra automorphism~\cite{Bagarello2022}, thereby leading to an apparent inequivalence between the Heisenberg and Schr\"{o}dinger representations. That is, quantum theory makes different predictions for the measurement with different representations.

		For instance, in general non-Hermitian systems, where $\ind{H}{S} (t) \neq \ind{H}{S}^\dagger (t)$, we have $U^\dagger (t, t_0) \neq U^{-1} (t, t_0)$. Hence, the commutation relations in Eqs.~\eqref{CommutationRelations1}-\eqref{CommutationRelations3} become time-dependent, e.g.,
		\begin{align}
			\begin{split}
				& \big[ q_{\text{\tiny H}i} (t), p_{\text{\tiny H}j} (t) \big]\\
				& = \big[ U^\dagger (t, t_0) q_{\text{\tiny S}i} (t) U (t, t_0) , U^\dagger (t, t_0) p_{\text{\tiny S}j} (t) U (t, t_0) \big]\\
				& \neq i \delta_{ij}.
			\end{split}
		\end{align}

		Additionally, since the Heisenberg operators $\ind{\mathcal{O}}{H} (t)$, as defined in Eq.~\eqref{HeisengbergAndSchroedingerOperator}, are no longer related to the corresponding Schr\"odinger operators $\ind{\mathcal{O}}{S} (t)$ by a simple similarity transformation, the measurement outcomes predicted by the Schr\"{o}dinger and Heisenberg representations can differ significantly.

		However, this apparent inequivalence between the two representations arises from the use of an inadequate quantum formalism. In the following, we demonstrate that these discrepancies are resolved within the metricized formalism \cite{Ju2019}.

	\section{The Schr\"{o}dinger Representation in the Metricized Formalism}

		To demonstrate that the inconsistencies between the Heisenberg and Schr\"{o}dinger representations can be resolved using the metricized formalism, we present the time-evolution operators for both the states and the dual states in the Schr\"{o}dinger representation.

		In the {\it metricized} formalism, the geometry of the Hilbert space must be accounted for when calculating the inner product between states, i.e.,
		\begin{equation}
			\SBraket{\psi_1 (t)}{\psi_2 (t)} = \Sbra{\psi_1 (t)} G (t) \Sket{\psi_2 (t)},
		\end{equation}
		where $G (t)$ is the metric of the Hilbert space, where the states and the dual-states are given by
		\begin{align}
			\SKet{\psi_2 (t)} \equiv \Sket{\psi_2 (t)} \text{ and } \SBra{\psi_1 (t)} \equiv \Sbra{\psi_1 (t)} G (t).
		\end{align}
		The time-evolution of states and the metric in the Schr\"{o}dinger representation are governed by:
		\begin{align}
			& \partial_t \SKet{\psi (t)} = - i \ind{H}{S} (t) \SKet{\psi (t)}, \label{SchroedingerEq}\\
			& \partial_t G (t) = i \left[ G (t) \ind{H}{S} (t) - \ind{H}{S}^\dagger (t) G (t) \right] \label{GEOM}.
		\end{align}
		It is worth mentioning that if $G(t_0)$ is Hermitian [$G^\dagger (t_0) = G (t_0)$], invertible, and positive-definite at a given time slice $t = t_0$, then $G(t)$ remains Hermitian, invertible, and positive-definite for all time~\cite{Mostafazadeh2004}.

		In the non-Hermitian regime, the metricized dual states are not simply the conventional ``Hermitian conjugate'' of the states. Consequently, the time-evolution operators for the states and their dual states require careful consideration.

		Here, we start with the states. For the remainder of the text, we place two ``mathematical nicety'' constraints on our Hamiltonian: Firstly, that $H(t)$ is bounded for all times $t \in \mathbb{R}$ and, secondly, that the map $H:t \to \mathcal{B}(\mathcal{H})$ is continuous with respect to the operator norm topology, where $\mathcal{B}(\mathcal{H})$ denotes the space of bounded linear operators on the Hilbert space $\mathcal{H}$. These constraints, which are automatically satisfied when $\mathcal{H}$ is finite-dimensional, guarantees the existence of a unique \textit{time-evolution operator}, $U_R(t,t_0)$, such that \cite[Thm. 5.1]{Pazy1983}
		\begin{equation}
			\SKet{\psi (t)} = \ind{U}{R} (t, t_0) \SKet{\psi (t_0)}.
		\end{equation}
		Together with Eq.~\eqref{SchroedingerEq}, the time-evolution operator satisfies
		\begin{equation}
			\partial_t \ind{U}{R} (t, t_0) = - i \ind{H}{S} (t) \ind{U}{R} (t, t_0), \label{RightEvolution}
		\end{equation}
		with $\ind{U}{R} (t_0,t_0) = \mathbbm{1}$.

		On the other hand, the time derivative of the dual state is determined not only by the Schr\"{o}dinger equation, but also by the metric equation given in Eq.~\eqref{GEOM}, as shown explicitly:
		\begin{align}
			\begin{split}
				\partial_t \SBra{\psi (t)} & = \partial_t \left[ \Sbra{\psi (t)} G (t) \right]\\
				& = i \left[ \Sbra{\psi (t)} \ind{H}{S} ^\dagger (t) \right] G (t)\\
				& \quad + i \Sbra{\psi (t)} \left[ G (t) \ind{H}{S} (t) - \ind{H}{S} ^\dagger (t) G (t) \right]\\
				& = i \Sbra{\psi (t)} G (t) \ind{H}{S} (t)\\
				& = i \SBra{\psi (t)} \ind{H}{S} (t).
			\end{split}
		\end{align}

		Hence, the time-evolution operator $\ind{U}{L} (t, t_0)$ for the dual state, namely,
		\begin{equation}
			\SBra{\psi (t)} = \SBra{\psi (t_0)} \ind{U}{L} (t, t_0),
		\end{equation}
		should satisfy
		\begin{align}
			\partial_t \ind{U}{L} (t, t_0) = i \ind{U}{L} (t, t_0) \ind{H}{S} (t), \label{LeftEvolution}
		\end{align}
		with $\ind{U}{L} (t_0,t_0) = \mathbbm{1}$.

		As a result, the time-evolution of the expectation value of an observable $\mathcal{O}$ in the Schr\"{o}dinger representation can be written as
		\begin{align}
			\begin{split}
				\left< \mathcal{O} (t) \right> & = \SBra{\psi (t)} \ind{\mathcal{O}}{S} (t) \SKet{\psi (t)}\\
				& = \SBra{\psi (t_0)} \ind{U}{L} (t, t_0) \ind{\mathcal{O}}{S} (t) \ind{U}{R} (t, t_0) \SKet{\psi (t_0)},
			\end{split} \label{SchroedingerExpectation}
		\end{align}
		where $\ind{\mathcal{O}}{S}$ is the corresponding operator for $\mathcal{O}$ in the Schr\"{o}dinger representation.

		By applying Eqs.~\eqref{RightEvolution} and \eqref{LeftEvolution}, we find
		\begin{align}
			& \partial_t \left[\ind{U}{L} (t, t_0) \ind{U}{R} (t, t_0)\right] = 0,\\
			& \partial_t \left[\ind{U}{R} (t, t_0) \ind{U}{L} (t, t_0)\right] = i \left[\ind{U}{R} (t, t_0) \ind{U}{L} (t, t_0), H(t)\right].
		\end{align}
		Together with the initial conditions $\ind{U}{R} (t_0, t_0) = \mathbbm{1} = \ind{U}{L} (t_0, t_0)$, it follows that
		\begin{align}
			& \ind{U}{L} (t, t_0) \ind{U}{R} (t, t_0) = \mathbbm{1} = \ind{U}{R} (t, t_0) \ind{U}{L} (t, t_0)\\
			\Rightarrow & \ind{U}{L} (t, t_0) = \ind{U}{R}^{-1} (t, t_0). \label{Inverse}
		\end{align}

		For simplicity, we omit the subscripts ``R'' and ``L,'' and define
		\begin{equation}
			U(t, t_0) \equiv \ind{U}{R} (t, t_0),
		\end{equation}
		such that
		\begin{equation}
			\ind{U}{L} (t, t_0) = U^{-1}(t, t_0).
		\end{equation}

		Not only is this property crucial for retaining the algebra automorphism property of time evolution, but it is also essential for maintaining consistency between the two representations, deriving the Heisenberg equation of motion, and validating the canonical commutation relations in the Heisenberg representation.

	\section{The Heisenberg representation in the Metricized Formalism}

		In the Heisenberg representation, the states and the dual state do not evolve with time, i.e.,
		\begin{align}
			& \HKet{\psi} = \SKet{\psi (t_0)},\\
			& \HBra{\psi} = \SBra{\psi (t_0)} = \Sbra{\psi (t_0)} G (t_0), \label{HeisenbergBra}
		\end{align}
		while all time-dependence is carried by the operators. Hence, the expectation value of the quantity $\mathcal{O}$ in the Heisenberg representation is given by
		\begin{align}
			\left< \mathcal{O} (t) \right> = \HBra{\psi} \ind{\mathcal{O}}{H} (t) \HKet{\psi}. \label{HeisenbergExpectation}
		\end{align}

		Comparing Eq.~\eqref{HeisenbergExpectation} with Eq.~\eqref{SchroedingerExpectation}, we find
		\begin{equation}
			\ind{\mathcal{O}}{H} (t) = U^{-1} (t, t_0) \ind{\mathcal{O}}{S} (t) U (t, t_0).\label{NHHRandSR}
		\end{equation}

		It is worth mentioning that, since $\ind{\mathcal{O}}{H} (t)$ is a similarity transformation of $\ind{\mathcal{O}}{S} (t)$, both $\ind{\mathcal{O}}{H} (t)$ and $\ind{\mathcal{O}}{S} (t)$ are isospectral, i.e., they share the same set of eigenvalues. Therefore, their possible measurement outcomes are identical.

		Next, we turn to the governing equation for the operators. Taking the time derivative of $\left< \mathcal{O} (t) \right>$ and obtain
		\begin{align}
			\frac{d}{dt} \left< \mathcal{O} (t) \right> = \HBra{\psi} \frac{d}{dt}\ind{\mathcal{O}}{H} (t) \HKet{\psi} \label{TimeDerivativeHeisenberg}
		\end{align}
		since $\HBra{\psi}$ and $\HKet{\psi}$ are time-independent. In contrast, in the Schr\"odinger representation, it holds
		\begin{align}
			\begin{split}
				\frac{d}{dt} \left< \mathcal{O} (t) \right> & = i \SBra{\psi (t)} \ind{H}{S} (t) \ind{\mathcal{O}}{S} (t)\SKet{\psi (t)}\\
				& \quad + \SBra{\psi (t)} \big[ \partial_t \ind{\mathcal{O}}{S} (t) \big] \SKet{\psi (t)}\\
				& \quad - i \SBra{\psi (t)} \ind{\mathcal{O}}{S} (t) \ind{H}{S} (t) \SKet{\psi (t)}.
			\end{split} \label{TimeDerivativeSchroedinger}
		\end{align}

		Since both Eqs.~\eqref{TimeDerivativeHeisenberg} and \eqref{TimeDerivativeSchroedinger} hold for any arbitrary state, we find that:
		\begin{align}
			& \SKet{\psi (t)} = U (t, t_0) \HKet{\psi},\\
			& \SBra{\psi (t)} = \HBra{\psi} U^{-1} (t, t_0),
		\end{align}
		imply
		\begin{align}
			\begin{split}
				\frac{d}{dt}\ind{\mathcal{O}}{H} (t) & = i U^{-1} (t, t_0) \ind{H}{S} (t) \ind{\mathcal{O}}{S} (t) U (t, t_0)\\
				& \quad - i U^{-1} (t, t_0) \ind{\mathcal{O}}{S} (t) \ind{H}{S} (t) U (t, t_0)\\
				& \quad + U^{-1} (t, t_0) \big[ \partial_t \ind{\mathcal{O}}{S} (t) \big] U (t, t_0).
			\end{split}
		\end{align}

		By inserting ${\mathbbm{1} = U^{-1} (t, t_0) U (t, t_0)}$ ${\big[\mathbbm{1} = U (t, t_0) U^{-1} (t, t_0) \big]}$ to the right (left) of $\ind{\mathcal{O}}{S} (t)$ in the first two terms of the above equation and rewriting all the operators in the Heisenberg representation, we arrive at
		\begin{align}
			\frac{d}{dt}\ind{\mathcal{O}}{H} (t) = i \big[ \ind{H}{H} (t), \ind{\mathcal{O}}{H} (t) \big] + \big[ \partial_t \mathcal{O} (t) \big]_\text{\tiny H}, \label{HeisenbergEOM}
		\end{align}
		where
		\begin{align}
			& \ind{H}{H} (t) = U^{-1} (t, t_0) \ind{H}{S} (t) U (t, t_0),\\
			& \big[ \partial_t \mathcal{O} (t) \big]_\text{\tiny H} = U^{-1} (t, t_0) \big[ \partial_t \ind{\mathcal{O}}{S} (t) \big] U (t, t_0).
		\end{align}

		Note that Eq.~\eqref{HeisenbergEOM} is, in fact, formally the same as the Heisenberg equation of motion. The only difference is that the metric must be considered in the dual state [Eq.~\eqref{HeisenbergBra}] as it is in the Schr\"{o}dinger representation.

		Moreover, Eq.~\eqref{NHHRandSR}, together with Eq.~\eqref{Inverse}, implies that the commutation relations in the Heisenberg representation are preserved, i.e.,
		\begin{align}
			& \mathcal{O}_{\text{\tiny S}[AB]} (t) \equiv \big[ \mathcal{O}_{\text{\tiny S}A} (t) , \mathcal{O}_{\text{\tiny S}B} (t)\big]\\
			\begin{split}
				& \Rightarrow \big[ \mathcal{O}_{\text{\tiny H}A} (t) , \mathcal{O}_{\text{\tiny H}B} (t) \big]\\
				& = \big[ U^{-1} (t, t_0) \mathcal{O}_{\text{\tiny S}A} (t) U (t, t_0), U^{-1} (t, t_0) \mathcal{O}_{\text{\tiny S}B} (t) U (t, t_0) \big]\\
				& = U^{-1} (t, t_0) \big[ \mathcal{O}_{\text{\tiny S}A} (t) , \mathcal{O}_{\text{\tiny S}B} (t) \big] U (t, t_0)\\
				& = U^{-1} (t, t_0) \mathcal{O}_{\text{\tiny S}[AB]} (t) U (t, t_0)\\
				& = \mathcal{O}_{\text{\tiny H}[AB]} (t).
			\end{split} \label{SchroedingerCanonicalCommutationRelation}
		\end{align}
		In other words, the commutation relations in both representations are consistent.

		Moreover, as in the Hermitian case, we examine the canonical commutation relations in the non-Hermitian regime. Let ${\lbrace (q_{\text{\tiny S}i}, p_{\text{\tiny S}i}) \rbrace}$ be the set of canonical conjugate quantity pairs. The canonical commutation relations in the Schr\"{o}dinger representation are:
		\begin{align}
			& \big[ q_{\text{\tiny S}i}, q_{\text{\tiny S}j} \big] = 0,\\
			& \big[ p_{\text{\tiny S}i}, p_{\text{\tiny S}j} \big] = 0,\\
			& \big[ q_{\text{\tiny S}i}, p_{\text{\tiny S}j} \big] = i \delta_{ij}.
		\end{align}
		Using Eq.~\eqref{SchroedingerCanonicalCommutationRelation}, the commutation relations in the Heisenberg representation become:
		\begin{align}
			& \big[ q_{\text{\tiny H}i} (t), q_{\text{\tiny H}j} (t) \big] = 0,\\
			& \big[ p_{\text{\tiny H}i} (t), p_{\text{\tiny H}j} (t) \big] = 0,\\
			& \big[ q_{\text{\tiny H}i} (t), p_{\text{\tiny H}j} (t) \big] = U^{-1} (t, t_0) i \delta_{ij} U (t, t_0) = i \delta_{ij}.
		\end{align}

		In other words, the commutation relations remain the same in the Heisenberg representation for closed quantum systems if the metricized formalism is applied.

	\section{Heisenberg-like Representation Based on the Generalized Vielbein Formalism}
		\begin{table*}[t]

			\renewcommand*{\arraystretch}{1.5}
			\begin{tabular}{| >{\centering\arraybackslash}m{0.18\textwidth} | >{\centering\arraybackslash}m{0.39\textwidth} | >{\centering\arraybackslash}m{0.39\textwidth} |}
				\hline
				& Metric & Vielbein\\
				\hline
				Metric and vielbein & $\displaystyle G(t) = \mathcal{E}^\dagger (t) \mathcal{E}(t)$ & $\mathcal{E} (t)$\\
				\hline
				States & $\SKet{\psi (t)} = \Sket{\psi (t)}$ & $\Vket{\psi (t)} = \mathcal{E} (t) \Sket{\psi(t)}$\\
				\hline
				Dual states & $\SBra{\psi (t)} = \Sbra{\psi (t)} G (t)$ & $\Vbra{\psi (t)} = \Sbra{\psi (t)} \mathcal{E}^\dagger (t)$\\
				\hline
				Operators & $\ind{\mathcal{O}}{S} (t)$ & $\ind{\mathcal{O}}{V} (t) = \mathcal{E} (t) \ind{\mathcal{O}}{S} (t) \mathcal{E}^{-1} (t)$\\
				\hline
				Inner product & $\SBraket{\phi (t)}{\psi (t)} = \Sbra{\phi (t)} G (t) \Sket{\psi (t)}$ & $\Vbraket{\phi (t)}{\psi (t)} = \Sbra{\phi (t)} \mathcal{E}^\dagger (t) \mathcal{E} (t) \Sket{\psi (t)}$\\
				\hline
				Governing equations & \begin{tabular}{c}
						$\partial_t \SKet{\psi (t)} = - i H (t) \SKet{\psi (t)}$,\\
						$\partial_t \SBra{\psi (t)} = i \SBra{\psi (t)} H (t)$,\\
						$\partial_t G (t) = i \left[ G (t) H (t) - H^\dagger (t) G (t) \right]$
					\end{tabular} & \begin{tabular}{c}
						$\partial_t \Vket{\psi (t)} = - i H_\flat (t) \Vket{\psi (t)}$,\\
						$\partial_t \Vbra{\psi (t)} = i \Vbra{\psi (t)} H_\flat (t)$,\\
						$H_\flat (t) = \mathcal{E} (t) H (t) \mathcal{E}^{-1} (t) + i \left[ \partial_t \mathcal{E} (t) \right] \mathcal{E}^{-1} (t)$
				\end{tabular}\\
				\hline
			\end{tabular}
			\caption{Comparing the representations described by the metric and the vielbein, it can be shown that $H_\flat(t)$ is a Hermitian time generator, i.e., $H_\flat^\dagger(t) = H_\flat(t)$. Therefore, the time evolution of states and the inner product between states in the vielbein formalism are identical to those in conventional Hermitian quantum mechanics, but the ``non-Hermiticity'' is transferred to the operators.}
			\label{Table:MetricAndVielbein}

		\end{table*}

		Besides the Schr\"{o}dinger and Heisenberg representations, there are many other representations that also describe quantum mechanics equivalently. One of these is the application of the generalized vielbein formalism~\cite{Ju2022}. In this formalism, the Hamiltonian is replaced by a Hermitian one, while the ``non-Hermiticity'' is carried by the operators.

		Specifically, the metric $G (t)$ can be decomposed into the generalized vielbein (abbreviated as ``vielbein''), namely,
		\begin{align}
			G (t) = \mathcal{E}^\dagger (t) \mathcal{E} (t),
		\end{align}
		where the vielbein satisfies the relation
		\begin{align}
			H_\flat (t) = \mathcal{E} (t) \ind{H}{S} (t) \mathcal{E}^{-1} (t) + i \left[\partial_t \mathcal{E} (t)\right] \mathcal{E}^{-1} (t), \label{VielbeinEq}
		\end{align}
		and $H_\flat (t)$ is an arbitrary Hermitized Hamiltonian that satisfies $H_\flat^\dagger (t) = H_\flat (t)$. The states, dual states, and operators are transformed by the vielbein as:
		\begin{align}
			& \Vket{\psi (t)} \equiv \mathcal{E} (t) \Sket{\psi (t)},\\
			& \Vbra{\psi (t)} \equiv \left[\Vket{\psi (t)}\right]^\dagger = \Sbra{\psi (t)} \mathcal{E}^\dagger (t),\\
			& \ind{\mathcal{O}}{V} (t) \equiv \mathcal{E} (t) \ind{\mathcal{O}}{S} (t) \mathcal{E}^{-1} (t),
		\end{align}
		 so that
		\begin{align}
			\begin{split}
				\SBra{\psi_1 (t)} \ind{\mathcal{O}}{S} (t) \SKet{\psi_2 (t)} & = \Sbra{\psi_1 (t)} G(t) \ind{\mathcal{O}}{S} (t) \Sket{\psi_2 (t)}\\
				& = \Vbra{\psi_1 (t)} \ind{\mathcal{O}}{V} (t) \Vket{\psi_2 (t)}{,}
			\end{split}
		\end{align}
		where the subscript ``V'' stands for ``vielbein.'' A comparison between the representations described by the metric and the vielbein can be found in Table~\ref{Table:MetricAndVielbein}.

		It is worth noting that the time evolution of the vielbein-transformed state is governed by
		\begin{align}
			\partial_t \Vket{\psi (t)} = - i H_\flat (t) \Vket{\psi (t)}. \label{VielbeinStateEq}
		\end{align}
		Since $H_\flat (t)$ is arbitrary, we can choose $H_\flat (t) = 0$ so that Eq.~\eqref{VielbeinEq} reduces to
		\begin{align}
			\partial_t \mathcal{E} (t) = i \mathcal{E} (t) \ind{H}{S} (t),
		\end{align}
		and Eq.~\eqref{VielbeinStateEq} renders states time-independent, i.e., $\Vket{\psi (t)} = \Vket{\psi (t_0)}$. Therefore, we can define:
		\begin{align}
			& \HLket{\psi} \equiv \Vket{\psi (t_0)} = \mathcal{E} (t_0) \Sket{\psi (t_0)},\\
			& \HLbra{\psi} \equiv \Vbra{\psi (t_0)} = \Sbra{\psi (t_0)} \mathcal{E}^\dagger (t_0),\\
			& \ind{\mathcal{O}}{HL} (t) \equiv \mathcal{E} (t) \ind{\mathcal{O}}{S} (t) \mathcal{E}^{-1} (t),
		\end{align}
		where the subscript ``HL'' stands for ``Heisenberg-like.'' Additionally, the time-evolution of operators becomes
		\begin{align}
			\frac{d}{dt}\ind{\mathcal{O}}{HL} (t) & = \frac{d}{dt} \left[\mathcal{E} (t)\ind{\mathcal{O}}{S} (t) \mathcal{E}^{-1} (t)\right]\\
			\begin{split}
				& = i \mathcal{E} (t) \ind{H}{S} (t) \ind{\mathcal{O}}{S} (t) \mathcal{E}^{-1} (t)\\
				& \quad - i \mathcal{E} (t) \ind{\mathcal{O}}{S} (t) \ind{H}{S} (t) \mathcal{E}^{-1} (t)\\
				& \quad + \mathcal{E} (t) \big[ \partial_t \ind{\mathcal{O}}{S} (t) \big] \mathcal{E}^{-1} (t)
			\end{split}\\
			\begin{split}
				& = i \ind{H}{HL} (t) \ind{\mathcal{O}}{HL} (t) - i \ind{\mathcal{O}}{HL} (t) \ind{H}{HL} (t)\\
				& \quad + \big[ \partial_t \mathcal{O} (t) \big]_\text{\tiny HL}
			\end{split}\\
			& = i \left[ \ind{H}{HL} (t), \ind{\mathcal{O}}{HL} (t) \right] + \big[ \partial_t \mathcal{O} (t) \big]_\text{\tiny HL},
		\end{align}
		where
		\begin{align}
			& \ind{H}{HL} = \mathcal{E} (t) \ind{H}{S} (t) \mathcal{E}^{-1} (t),\\
			& \big[ \partial_t \mathcal{O} (t) \big]_\text{\tiny HL} = \mathcal{E} (t) \left[\partial_t \ind{\mathcal{O}}{S} (t)\right] \mathcal{E}^{-1} (t).
		\end{align}

		\begin{table*}
			\renewcommand{\arraystretch}{1.5}
			\begin{tabular}{|c|c|c|}
				\hline
				& Heisenberg representation & Heisenberg-like representation\\
				\hline
				Time evolution & \begin{tabular}{c}$\displaystyle \partial_t U(t, t_0) = - i \ind{H}{S} (t) U(t, t_0)$\\$U(t_0, t_0) = \mathbbm{1}$\end{tabular} & \begin{tabular}{c}$\displaystyle \partial_t \mathcal{E} (t) = i \mathcal{E} (t) \ind{H}{S} (t)$\\$\displaystyle G (t_0) = \mathcal{E}^\dagger (t_0) \mathcal{E} (t_0)$\end{tabular}\\
				\hline
				Observables & $\displaystyle \ind{\mathcal{O}}{H}(t) = U(t, t_0) \ind{\mathcal{O}}{S}(t) U^{-1}(t, t_0)$ & $\displaystyle \ind{\mathcal{O}}{HL}(t) = \mathcal{E}(t) \ind{\mathcal{O}}{S}(t) \mathcal{E}^{-1} (t)$\\
				\hline
				Operators & $\displaystyle \frac{d}{dt}\ind{\mathcal{O}}{H}(t) = i \big[ \ind{H}{H}(t), \ind{\mathcal{O}}{H}(t) \big] + \big[ \partial_t \mathcal{O} (t) \big]_\text{\tiny H}$ & $\displaystyle \frac{d}{dt}\ind{\mathcal{O}}{HL}(t) = i \big[ \ind{H}{HL}(t), \ind{\mathcal{O}}{HL}(t) \big] + \big[ \partial_t \mathcal{O} (t) \big]_\text{\tiny HL}$ \vphantom{$\Bigg[$}\\
				\hline
				States & $\displaystyle \HKet{\psi} = \Hket{\psi} = \Sket{\psi(t_0)}$ & $\displaystyle \HLket{\psi} = \mathcal{E} (t_0) \Sket{\psi(t_0)}$\\
				\hline
				Dual states & $\displaystyle \HBra{\psi} = \SBra{\psi(t_0)} = \Sbra{\psi(t_0)} G(t_0)$ & $\displaystyle \HLbra{\psi} = \Sbra{\psi(t_0)} \mathcal{E}^\dagger (t_0) = \left(\HLket{\psi}\right)^\dagger$\\
				\hline
			\end{tabular}
			\caption{Comparing the Heisenberg and Heisenberg-like representations. Note that $\mathcal{E}^{-1} (t)$ plays a similar role to $U(t, t_0)$; however, unlike $U(t, t_0)$, $\mathcal{E} (t = t_0) \neq \mathbbm{1}$ in general. As a consequence, the states undergo a linear transformation such that their Hermitian conjugates are identical to their dual states.}
			\label{Table:HvsHL}
		\end{table*}

		In other words, choosing the vielbein with ${H_\flat (t) = 0}$ renders the states time-independent and the operators subject to the standard Heisenberg equation. However, unlike the standard Heisenberg representation, where only the dual states carry the information about the metric of the Hilbert space, the generalized-vielbein-formalism-induced Heisenberg-like representation distributes the metric information across both the states and the dual states (see Table~\ref{Table:HvsHL}). As a consequence, the relation between the states and the dual states in the Heisenberg-like representation is analogous to the standard Hermitian conjugate, namely,
		\begin{align}
			\begin{split}
				\HLbra{\psi} & = \Sbra{\psi (t_0)} \mathcal{E}^\dagger (t_0) = \left[\mathcal{E} (t_0) \Sket{\psi (t_0)}\right]^\dagger\\
				& = \left(\HLket{\psi}\right)^\dagger.
			\end{split}
		\end{align}

		Therefore, although the states and dual states are time-independent and the operators are formally governed by the same equation of motion in both Heisenberg and Heisenberg-like representations, the metric dependence is much less apparent in the Heisenberg-like representation.

	\section{Conclusions}

		The equivalence between the Schr\"{o}dinger and Heisenberg representations in quantum mechanics is fundamental, both conceptually and for potential applications. Any inconsistency between the two would lead to various issues, such as discrepancies in predicting measurement outcomes, violations of the uncertainty principle, and the breakdown of quantization schemes. However, in the non-Hermitian regime, the conventional transformation between these representations leads to a discrepancy.

		To establish the equivalence between the two representations, we turn to the metricized formalism, which incorporates the geometry of the Hilbert space bundle. This formalism naturally generalizes conventional quantum mechanics — extending its applicability to the non-Hermitian regime while offering deeper insights and serving as a powerful analytical framework.

		By applying the metricized formalism, we derive the relationship between the two representation, which reduces to the conventional one in the Hermitian regime. Moreover, we demonstrate their equivalence by showing that the predicted measurement outcomes are identical, the commutation relations are consistent, and the canonical commutation relations remain unaltered. Additionally, we provide the Heisenberg equation of motion for operators in the non-Hermitian regime.

		It is worth mentioning that not only have the discussions been extended beyond the quasi-Hermitian regime (i.e., the regime with a dynamical metric), but they can also be applied to cases where the Hamiltonian of the system is time-dependent.

		In addition to these two representations, we also introduce a Heisenberg-like representation via the generalized vielbein formalism, which renders the metric information implicit. In the Heisenberg representation, the metric is encoded exclusively in the dual states, whereas in the Heisenberg-like representation, the metric information is distributed across both the states and the dual states. Nevertheless, in both representations, all states and dual states remain time-independent, and the operators satisfy the Heisenberg equation of motion.

		Consequently, this work establishes a physically consistent transformation between the Schr\"{o}dinger, Heisenberg, and Heisenberg-like representations of quantum mechanics through the metricized formalism.

	\begin{acknowledgments}
		C.Y.J. is partially supported by the National Science and Technology Council (NSTC) through Grant No. NSTC 112-2112-M-110-013-MY3. A.M. was supported by the Polish National Science Centre (NCN) under the Maestro Grant No. DEC-2019/34/A/ST2/00081. J.B. is supported by the Basque government through the BERC 2022-2025 program. G.Y.C is partially supported by the NSTC through Grants No. 113-2112-M-005-008 and 113-2123-M-006-001. F.N. is supported in part by: the Japan Science and Technology Agency (JST) [via the CREST Quantum Frontiers program Grant No. JPMJCR24I2, the Quantum Leap Flagship Program (Q-LEAP), and the Moonshot R\&D Grant Number JPMJMS2061], and the Office of Naval Research (ONR) Global (via Grant No. N62909-23-1-2074).
	\end{acknowledgments}

	\bibliography{References}

\end{document}